# Demonstration of Si based InAs/GaSb type-II superlattice p-i-n photodetector

Zhuo Deng*, Daqian Guo*, Claudia González Burguete, Jian Huang, Zongheng Xie, Huiyun Liu, *Senior Member, IEEE,* Jiang Wu, *Member, IEEE,* Baile Chen, *Member, IEEE*

***Abstract*—In this paper, mid-wave infrared photodetection based on an InAs/GaSb type-II superlattice p-i-n photodetector grown directly on Si substrate is demonstrated and characterized. Excitation power dependence on integrated intensity from the photoluminescence measurements reveals a power coefficient of $P\sim I^{0.74}$, indicating that defects related process is playing an important role in the predominant recombination channel for photogenerated carriers. At 70 K, the device exhibits a dark current density of 2.3 A/cm$^2$ under -0.1 V bias. Arrhenius analysis of dark current shows activation energies much less than half of the active layer bandgap, which suggests that the device is mainly limited by surface leakage and defect-assisted tunneling, consistent with the photoluminescence analysis. The detector shows 50% cutoff wavelength at ~5.5 µm at 70 K under bias of -0.1 V. The corresponding peak responsivity and specific detectivity are 1.2 A/W and $1.3\times10^9$ cm·Hz$^{1/2}$/W, respectively. Based on these optoelectronics characterization results, reduction of defects by optimizing the III/V-Si interface, and suppression of surface leakage channels are argued to be the main factors for performance improvement in this Si-based T2SL detector towards low cost, large-format MWIR detection system on Si photonics platform.**

## I. Introduction

Mid-wave infrared (MWIR) photodetection based on InAs/GaSb type-II superlattice (T2SL) has recently generated extensive interests in both civil and military areas such as environmental monitoring, medical treatment and target imaging [1]. Compared with the commercially predominating HgCdTe (MCT) detectors which suffer from composition nonuniformity and high fabrication cost, the T2SL system enjoys several important advantages, including tunable detection wavelength by modifying the active layer thickness and composition, [2], lower Auger recombination rate [3] and lower tunneling dark current [4]. These figures of merits have enabled the InAs/GaSb T2SL to become the leading candidate for fabrication of high-performance focal plane arrays (FPAs) to accommodate the market demand for the technologically important MWIR detection range [5]. Nevertheless, one of the significant obstacles which impede the development of FPAs based on T2SL towards larger format, is the lack of low-cost substrates with satisfactory crystalline quality. Recent demonstration of III-V semiconductors grown on GaAs [6] and Si [7, 8] substrates have paved the way for direct integration of T2SL detectors on these cost effective non-native substrates. In particular, monolithic integration of T2SL structure on Si is of utmost interest, since it offers great potential for direct integration of MWIR detection modules with other active components on a single Si wafer which enables a complete large-scale Si photonic circuit system. However due to the large lattice mismatch between Si and T2SL (~14%), growth of high-performance T2SL photodetectors on Si is very challenging and many groups have devoted into solving this problem. For instance, we previously combined the interfacial misfit array technique and the AlSb/GaSb dislocation filter superlattice to generate the first report on direct growth of InAs/GaSb T2SL on Si by molecular beam epitaxy (MBE) [9]. Recently Durlin et al.[10] have employed the GaSb buffer layer on Si grown by metal organic chemical vapor deposition (MOCVD) to grow a barrier InAs/InAsSb T2SL detector which achieved cutoff wavelength of ~5 µm. Soon after Delli et al. [11] have demonstrated a similar barrier detector structure with cutoff of ~5.5 µm using the interfacial misfit array and the AlSb/GaSb dislocation filter layer to grow on Si. These results have already achieved good performances at cryogenic temperature, although further improvements have to be carried out in order to compete with the T2SL grown on native substrate.

In this study, we report the comprehensive electrical and optical characterization of an InAs/GaSb type-II superlattice p-i-n photodetector directly grown on Si substrate by MBE. The device covers a MWIR detection range from 3 to 6 µm, with a quantum

This work was supported by the Shanghai Sailing Program under Grant 17YF1429300, the ShanghaiTech University startup funding under Grant F-0203-16-002, the UK EPSRC First Grant EP/R006172/1 and UK DSTL grant DSTLX-1000107901. (*Corresponding authors: Jiang Wu and Baile Chen.*)

Z. Deng, J. Huang, Z. Xie and B. Chen are with the Optoelectronic Device Laboratory, School of Information Science and Technology, ShanghaiTech University, Shanghai 201210, Peoples R China (e-mail: dengzhuo@shanghaitech.edu.cn; huangjian@shanghaitech.edu.cn; xiezh@shanghaitech.edu.cn; chenbl@shanghaitech.edu.cn ).

D. Guo, C. Burguete, H. Liu and J. Wu are with the Department of Electronic and Electrical Engineering, University College London, London WC1E 7JE, United Kingdom (e-mail: daqian.guo.15@ucl.ac.uk; claudia.burguete.16@ucl.ac.uk; huiyun.liu@ucl.ac.uk; jiang.wu@ucl.ac.uk ).

J. Wu are currently with Institute of Fundamental and Frontier Sciences, University of Electronic Science and Technology of China, Chengdu 610054 Sichuan, Peoples Republic of China

* Z. Deng and D. Guo contribute equally to this work.

efficiency of 27.6% at 70 K and 5 μm under -0.1 V. The corresponding specific detectivity of the device is in the order of $10^9$ cm·Hz$^{1/2}$/W. Key device factors for future performance optimization are also addressed.

## II. Experimental Details

Schematic structure of the InAs/GaSb T2SL p-i-n detector grown on Si is shown in Fig. 1 (a). The layers were directly grown by molecular beam epitaxy on the Si (110) substrate with a 4° misorientation towards [011] direction in order to suppress the formation of anti-phase domains. Details of the growth processes have been reported in our previous work [9]. A low temperature AlSb nucleation layer with 10 nm thickness was first deposited on the Si substrate. Then a two-period buffer layer formed by 10 monolayer (ML) AlSb/10 ML GaSb superlattice (100 nm) and 500 nm of GaSb spacer was grown. The active layers consisted of 10 ML InAs/10 ML GaSb with 1ML of InSb inserted in between for strain balance. Standard thicknesses of 500 nm *n*-region/2 μm non-intentionally-doped (n.i.d) absorption region/500 nm *p*-region were grown for the p-i-n design. The device was capped with a 50 nm heavily $p^+$-doped GaSb contact layer. Based on the full-width-at-half-maximum (FWHM) of the GaSb (004) diffraction peak measured from the XRD (~284.9 arcsec) in Ref. [9], the surface dislocation density ($\rho_d$) of the T2SL wafer grown on Si is estimated to be ~$2.6\times10^8$ cm$^{-2}$, according to the equation $\rho_d$ = (FWHM/2$B$)$^2$, where $B = 0.431\times10^{-7}$ cm is the Burgers vector [12]. It should be noted that the dislocation density in the active region is much lower than the estimated value here, because the highly defective GaSb buffers also contribute to the FWHM of the GaSb (004) peak due to the deep penetration of x-ray.

The wafer used for the device fabrication in this work was equivalent to the samples grown in Ref. [9]. After the material growth, the sample was processed into mesa-isolated devices with various circular diameters ranging from 70 to 350 μm using standard contact-mode UV photolithography. Citric acid-based wet chemical etching with $C_6H_8O_7$:$H_3PO_4$:$H_2O_2$:$H_2O$ = 1:1:4:16 solution was used to define the mesa and the sidewalls were passivated by SU-8. Finally, metal contacts of Ti/Pt/Au (50 nm/50nm/300 nm) were deposited as the top and bottom contacts by e-beam evaporation and lift-off technique. No anti-reflection layers were coated on the device.

Optical properties of the T2SL device were probed by excitation power dependent photoluminescence (PL) at temperature of 10 K. The sample was mounted in a liquid helium closed-cycle cryostat and excited by a 532 nm laser with a beam diameter of ~1 mm. The pump power was varied from 10 to 600 mW (corresponding spectral density: 1.27 to 76.4 W/cm$^2$). PL signals were chopped and detected by a monochromator equipped with an InSb detector at liquid nitrogen temperature. The detected signals were then collected and analyzed by a lock-in amplifier. For electrical and photoresponse characterization, the sample was cleaved and a test device with 130 μm diameter was wire-bonded with a TO header. The chip was loaded into a variable-temperature cryostat with radiation shield installed and all surfaces of the cryostat were covered by aluminum foils. A semiconductor parameter analyzer was used to measure the dark current-voltage characteristics. The chip was remained in the cryostat for the spectral response measurement. A Fourier Transform Infrared Spectrometer in normal incidence setup and a low-noise current preamplifier were used to bias the chip and record the spectra. To obtain the responsivity, the chip was illuminated by a standard blackbody source operating at 700°C. The same current preamplifier was used to provide bias and measure the photocurrent simultaneously. The signal was then recorded by a Fast-Fourier-Transform network analyzer.

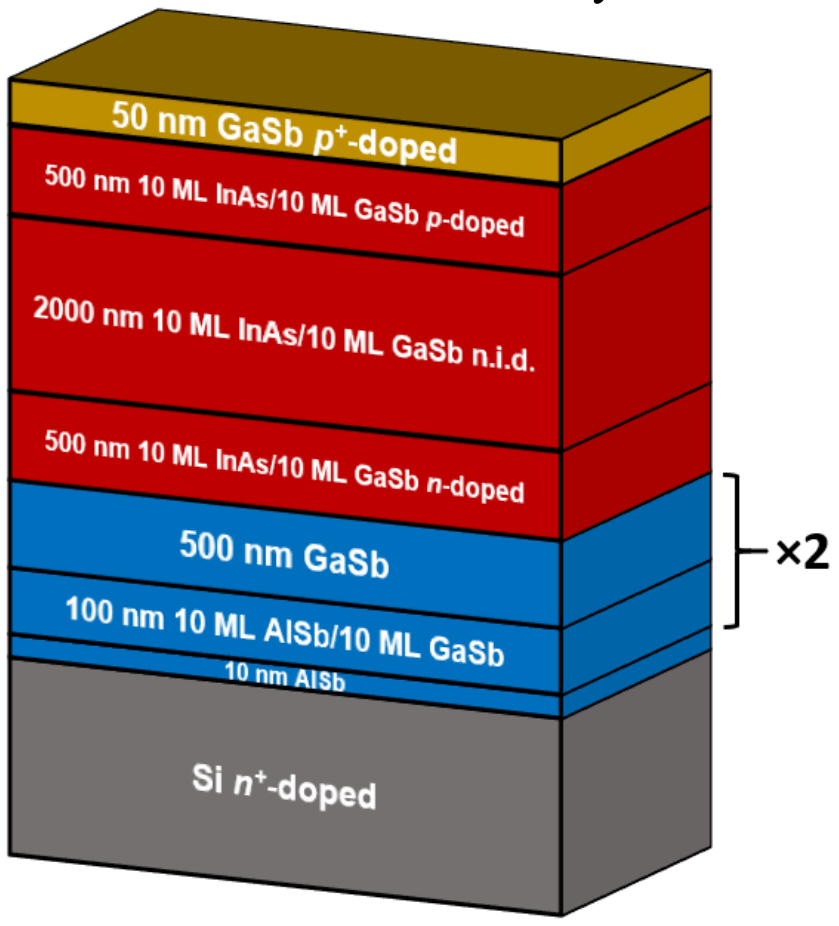

Fig. 1. Schematic diagram of the T2SL detector structure grown on Si substrate.

## III. Results and Discussions

Excitation power dependent PL spectra measured at 10 K from the T2SL as-grown wafer on Si are plotted in Fig. 2. A clear blue-shift in PL energy with increasing laser pump power is clearly observed. By computing the integrated PL intensity at different laser power, the dependence between the two quantities can be modeled by the power law relation: [13]

$$P_{Laser} = A_{SRH} I_{PL}^{1/2} + B_{RR} I_{PL} + C_{Auger} I_{PL}^{3/2} \qquad (1)$$

where $P_{Laser}$ is the laser pump power, $I_{PL}$ is the integrated PL intensity, and the leading coefficients $A_{SRH}$, $B_{RR}$ and $C_{Auger}$ are associated with the Shockley-Read-Hall (SRH), radiative recombination and Auger recombination channel respectively. As a result, by fitting the data using Eq. 1 one could unveil the signature of dominant carrier recombination process in the active layer from the power value. The inset of Fig. 2 shows the laser power as a function of PL integrated intensity. The best fit is obtained with a power relation of $P\sim I^{0.74\pm0.06}$, which lies between the theoretical value for the SRH ($P\sim I^{0.5}$) and radiative process ($P\sim I^{1}$). This result suggests that in this T2SL wafer grown on Si the trapping of carriers through defect levels in the bandgap plays an important role in the radiative recombination of photogenerated electrons and holes. Indeed, from the structural investigation by AFM and XRD characterization in our previous work [9], the crystal quality of the T2SL sample grown on Si suffers from the large number of the 60° misfits formation at the AlSb/Si interface. These misfits could facilitate the propagation of threading dislocation into the active layer, which generate efficient nonradiative recombination centers for photogenerated carriers, as revealed from the power law relation above.

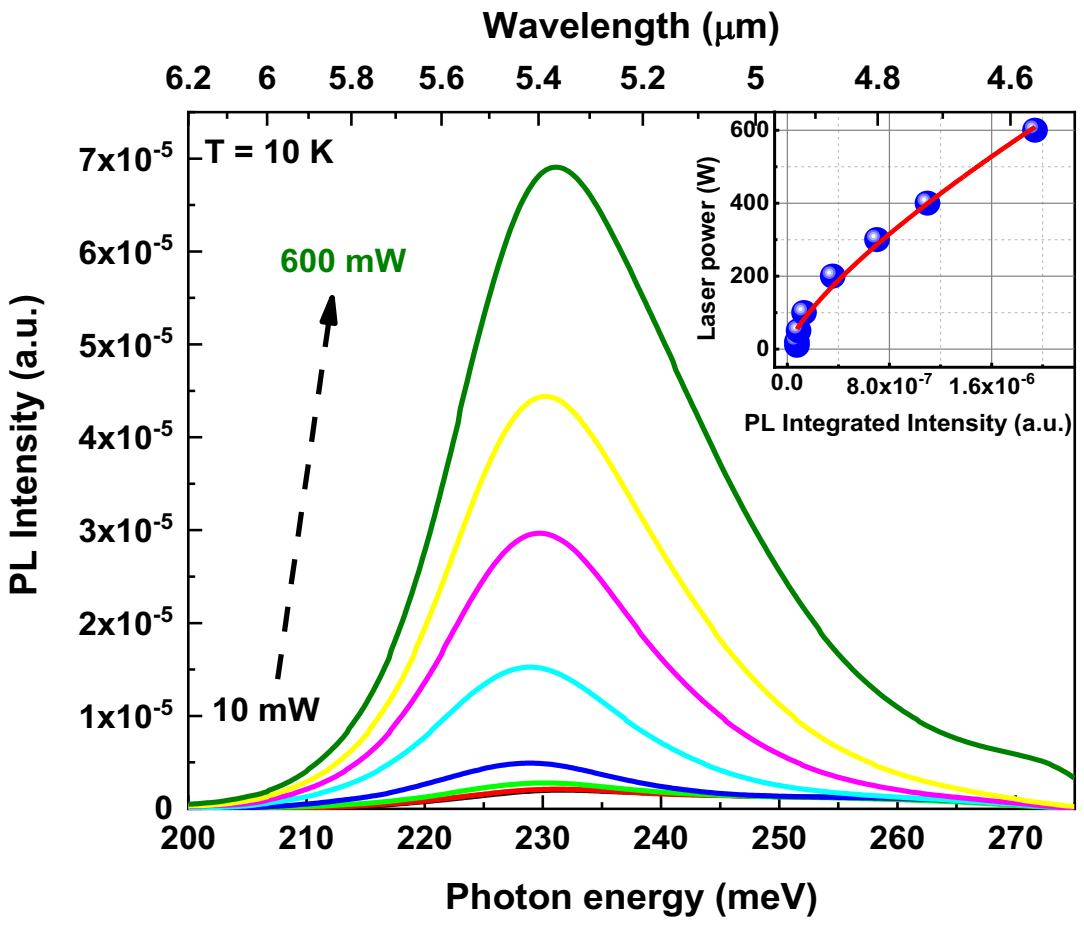

Fig. 2. Photoluminescence spectra measured at different laser pump power from the T2SL as-grown wafer on Si at 10 K. The inset plots the PL integrated intensity against the laser pump power, and the data are fitted by the power law model (Eq. 1) as indicated by the solid line.

Dark current density-voltage characteristic curves measured from the T2SL detector grown on Si at different temperatures are shown in Fig. 3 (a). The test device has a circular mesa diameter of 130 μm. At 50 K, the dark current density ($J_d$) of the device is found to be 2.3 A/cm$^2$ under -0.1 V bias, and it increases to 16.1 A/cm$^2$ at room temperature. Fig. 3 (b) shows the temperature dependence of differential-resistance-area product ($RA$) evaluated from the $J_d$-$V$ curves. The $RA$ under zero bias ($R_0A$) at 50 K is found to be $8.6\times10^{-2}$ $\Omega\cdot$cm$^2$. To further investigate the dark current generation mechanism, Arrhenius plots under -0.1 V bias are shown in Fig. 3 (c). Linear fits for the low temperature range (30-150 K) yields an activation energy ($E_{a1}$) of ~9.4 meV, whereas for the high temperature regime (170-300 K) the activation energy ($E_{a2}$) increases to ~31.3 meV. Both values are significantly smaller than half of the T2SL active layer bandgap energy, as shown by the PL spectra and the cutoff of spectral response (see Fig. 4). To gain further insight of the carrier transport mechanism, the unbonded processed wafer was loaded in a low-temperature probe station maintained at 77 K. The dark current of devices with different circular diameters (70 - 350 μm) were measured, and the results are shown in Fig. 3 (d). A good linear fit can be obtained from the data plot, despite a small extent of device inhomogeneity which may be resulted from growth and/or fabrication process. This suggests that at low temperature regime the dark current in these T2SL detectors are dominated by surface leakage, which is in agreement with the very small activation energy ($E_{a1}$ ~ 9.4 meV) obtained from the Arrhenius analysis. At higher temperature region the activation energy increases by a factor of ~3, which may be attributed to additional contribution from defect-assisted tunneling. Therefore, besides the needs for the reduction of dislocation (~$10^8$ cm$^{-2}$) from optimizing the growth condition, surface conductive channels of the device should be further suppressed by developing more effective sidewalls passivation technique in our future works.

Fig. 4 presents the absolute responsivity of the T2SL detector grown on Si measured at various temperatures under -0.1 V. At 70 K, the 50% cutoff wavelength is found at ~5.5 μm, and the peak responsivity is ~1.2 A/W. This value improves to ~1.7 A/W when temperature decreases to 50 K. On the other hand, the cutoff wavelength extends to ~6 μm at 130 K and the responsivity drops to ~0.1 A/W. The corresponding external quantum efficiency (EQE) of the T2SL device evaluated at different temperatures under -0.1 V are plotted in Fig. 5 (a). At 50 K and 5 μm the device exhibits an EQE of 44.6% under -0.1 V, and it reduces to 27.6% at 70 K, as extracted and shown in Fig. 5 (b). Further raise of temperature leads to a rapid decrease in EQE. At 130 K the device EQE falls to ~2%. We attribute this temperature induced drop in EQE to the increased trapping rate of dislocation centers in the T2SL absorption region at higher temperature, which reduces the diffusion length of photogenerated carrier and hence the extracted photocurrent. Similarly, the EQE exhibits monotonic decrease with increasing temperature when the device is operating at zero-bias mode, as can be seen in Fig. 5 (b). The slight increase from 110 K to 130 K is due the undulation of weak signal measured under zero bias. This indicates that the carrier collection efficiency is very low when the device is operating under zero bias (EQE < 1%), where most of the photogenerated carriers are captured by the defects. A small bias of -0.05 V is sufficient to fully extract the carriers which leads to two orders of magnitude enhancement in EQE.

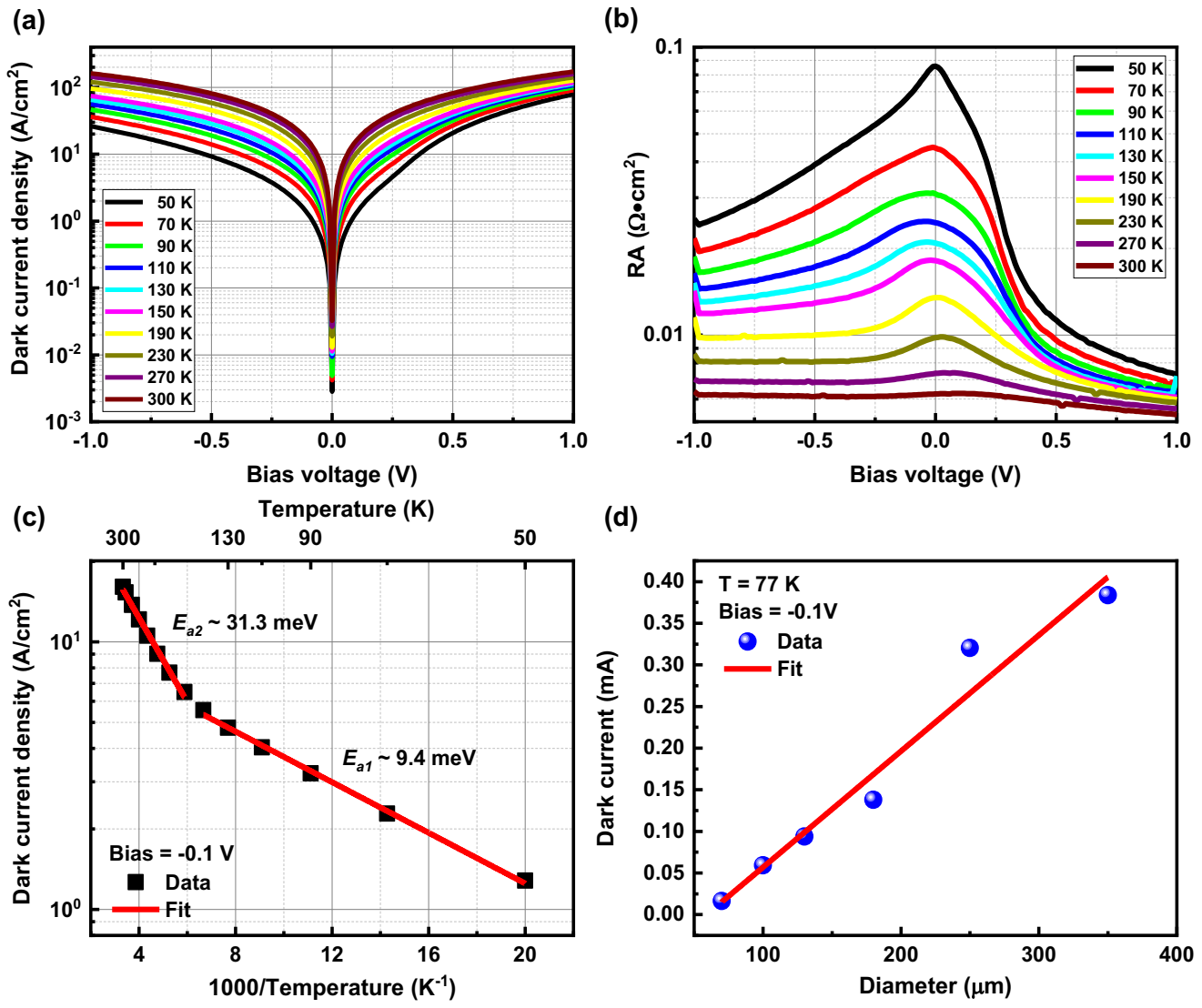

Fig. 3 (a) Dark current density-voltage characteristic of the T2SL detector grown on Si; (b) Differential-resistance-area product (*RA*) versus bias voltage at different temperatures; (c) Arrhenius plot and the linear fits of the dark current density under -0.1 V bias; (d) Dark current of the T2SL detectors as a function of circular diameter measured under -0.1 V and at 77 K.

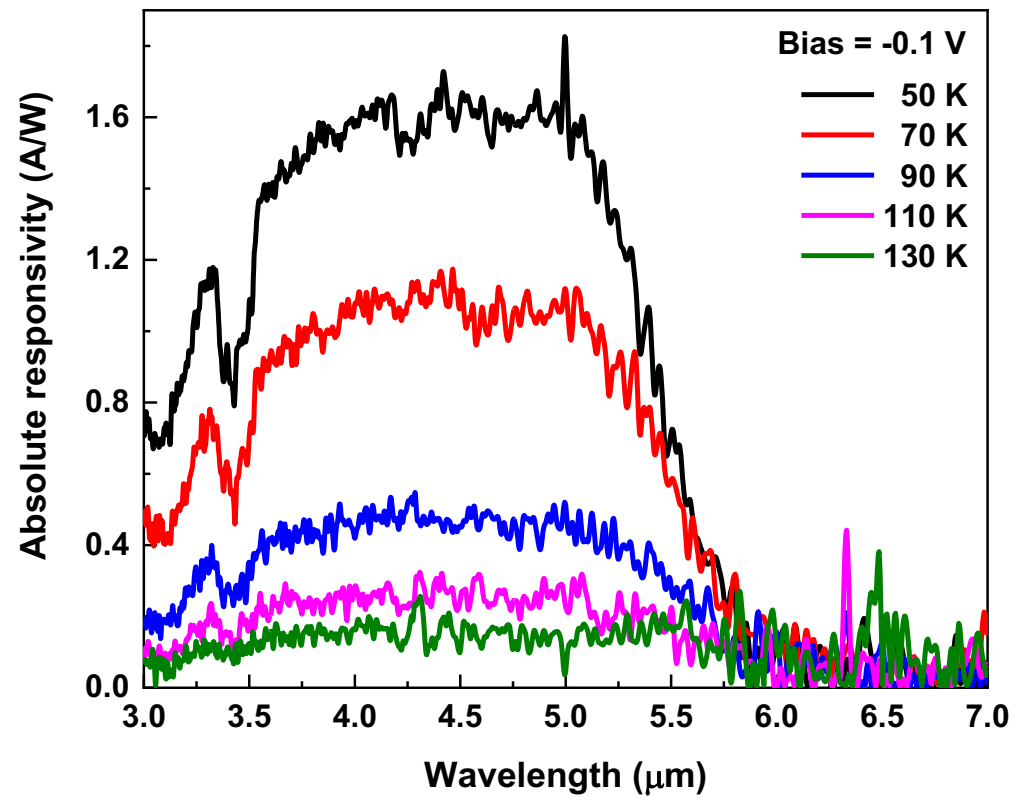

Fig. 4. Absolute responsivity of the T2SL detector grown on Si as a function of temperature under -0.1 V.

Under bias operation the detector noise is limited by both thermal and shot noise component. Thus the specific detectivity $D*$ can be calculated by:

$$D^* = \frac{R}{\sqrt{\frac{4k_BT}{R_0A} + 2qJ_d}} \tag{2}$$

where $R$ is the responsivity, $k_B$ is Boltzmann's constant, $T$ is the temperature of the device, $R_0A$ is the differential-resistance-area-product under zero bias, $q$ is the electron charge and $J_d$ is the dark current density under bias. Fig. 6 depicts the calculated $D*$ of the T2SL detector grown on Si versus temperature. At 70 K, a peak $D*$ of $1.3\times10^9$ cm·Hz$^{1/2}$/W is obtained, and it drops by an order of magnitude to ~$1.5\times10^8$ cm·Hz$^{1/2}$/W at 130 K. These values are lower than those reported recently in an Ga-free InAs/InAsSb T2SL nBn detector grown on Si substrate with similar operating wavelength [11]. In that structure five sets of similar AlSb/GaSb superlattice was used as dislocation filter buffer, which resulted in an order of magnitude lower dislocation density (~$3\times10^7$ cm$^{-2}$) compared with our current design. Thus, combining the above optoelectronic characterization results, further improvement of the current Si-based T2SL photodetector with simple p-i-n structure can be expected by redesigning the AlSb/GaSb dislocation filter superlattice (e.g., increase the iteration) to reduce the dislocation density in the active region, while still enjoying the design and growth simplicity of p-i-n structure.

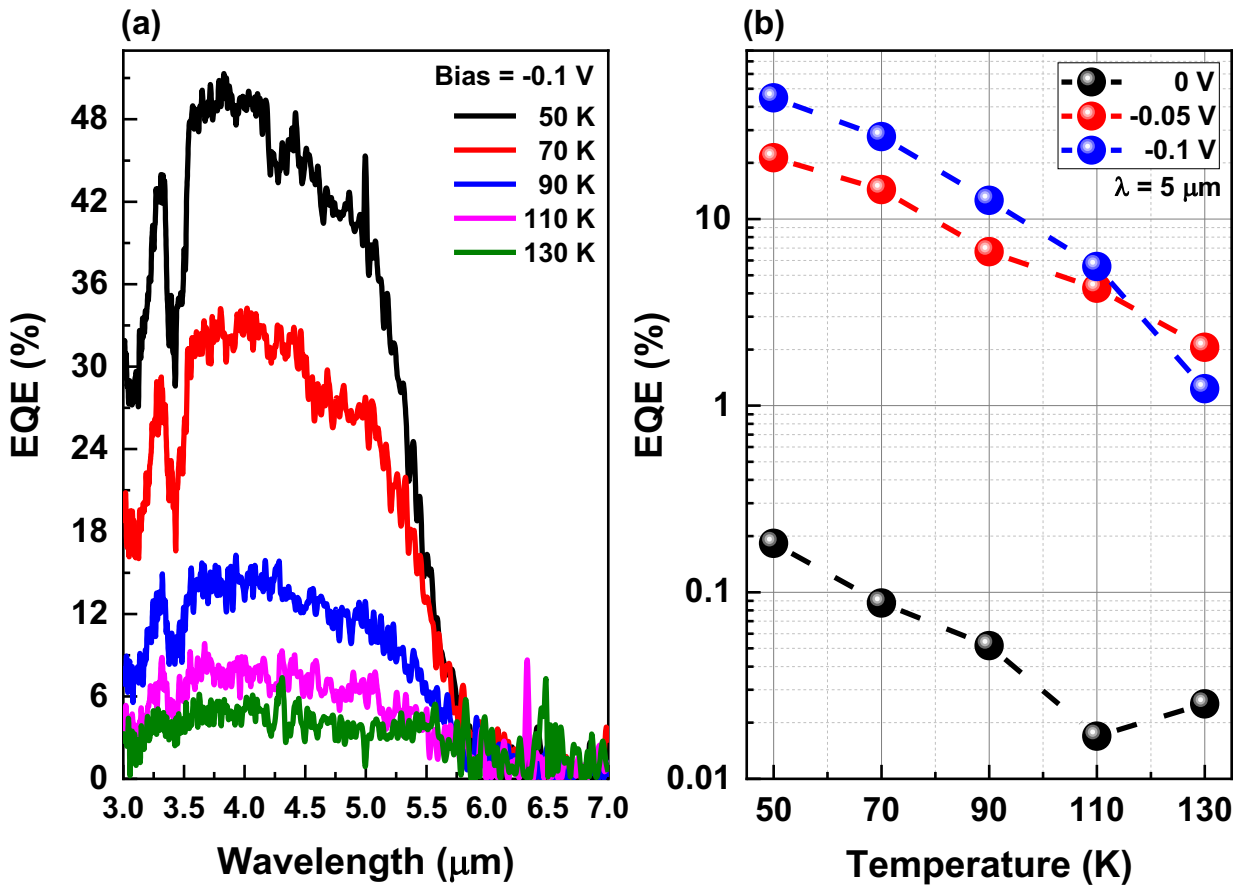

Fig. 5. External quantum efficiency (EQE) of the T2SL detector grown on Si (a) at different temperatures under -0.1 V; (b) Temperature dependence of EQE at 5 μm under three different bias voltages.

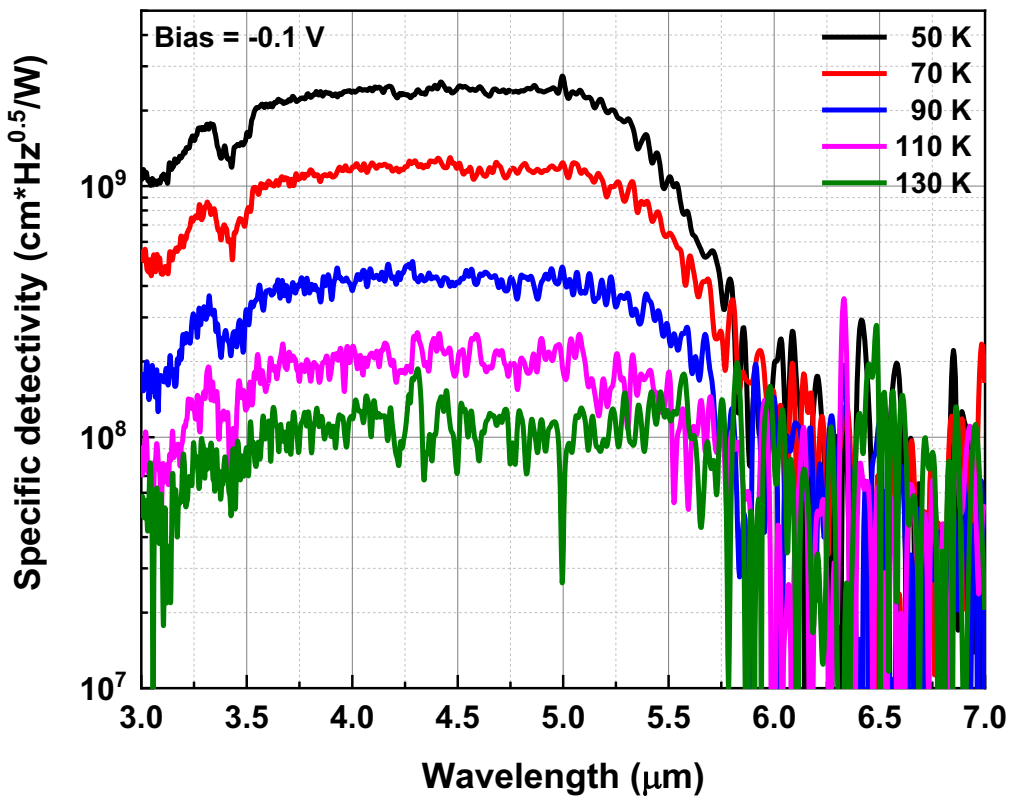

Fig. 6. Specific detectivity ($D^*$) of the T2SL detector grown on Si versus temperature under -0.1 V.

## IV. Conclusion

In conclusion, an InAs/GaSb T2SL p-i-n photodetector has been integrated directly on Si and the optoelectronics properties have been extensively characterized. Excitation dependent PL measured at low temperature suggests that the photogenerated carriers in the active regions are mainly limited by the SRH recombination channel related to dislocations, which is also evidenced by the high dark current density and small activation energy. At 70 K, the device shows 50% cutoff wavelength at ~5.5 μm under -0.1 V, and the corresponding peak responsivity and specific detectivity are 1.2 A/W and $1.3\times10^9$ cm·Hz$^{1/2}$/W, respectively. Based on these results, further performance improvement of this Si-based type-II superlattice detector should be focused on reduction of dislocation density in the active layer as well as suppression of the large surface leakage to increase the device operating temperature.